\documentclass[showpacs,preprint]{revtex4}
\usepackage{graphicx}
\usepackage{subfigure}
\usepackage{amssymb}
\usepackage{amsmath}
\usepackage{bm}
\usepackage{latexsym}
\usepackage{hyperref}

\begin{document}

\begin{titlepage}
\title
{Non-markovian dynamics of double quantum dot charge qubit with static bias}
\author{Xiufeng Cao\footnote{Email: cxf@sjtu.edu.cn}, Hang Zheng}

\address{Department of Physics, Shanghai Jiaotong University,
Shanghai 200240, People's Republic of China}

\begin{abstract}
The dynamics of charge qubit in double quantum dot coupled to
phonons is investigated theoretically. The static bias is
considered. By means of the perturbation approach based on unitary
transformations, the dynamical tunneling current is obtained
explicitly. The biased system displays broken symmetry and a
significantly larger coherence-incoherence transition critical point
$\alpha _{c}$. We also analyzed the decoherence induced by
piezoelectric coupling phonons in detail. The results show that
reducing the coupling
 between system and bath make coherence
frequency increase and coherence time prolong. To maintain quantum
coherence, applying static bias also is a good means.
\\


\end{abstract}

 \maketitle

\address{Department of Physics, Shanghai Jiaotong University,
Shanghai 200240, People's Republic of China}

\end{titlepage}

1. Introduction

Since the discovery that quantum algorithms can solve certain computational
problems much more efficiently than classical ones\cite{1}, attention has
been devoted to the physical implementation of quantum computation.
Nanofabrication technology now allow us to design artificial atoms (quantum
dots) and \ molecules (coupled quantum dots), in which atomic (molecular)
-like electronic states can be controlled with external gate voltages\cite%
{2,3,4}. As the system always interacts with its environment, quantum
decoherence in the system usually is the most serious obstacle to produce
efficient quantum circuits\cite{5,6,7}. For the reason, a detailed
understanding of quantum decoherence in open system and implementing
sufficiently high number of coherent manipulation within the characteristic
coherence time of qubits are crucial for future actual implementation of
quantum nanostructures to quantum information technology.

To build a quantum computer, the first step is the realization of the basic
device units for quantum information processing called quantum bit (qubit).
Within the last decade, various schemes have been proposed and many of them
have even been realized, such as superconducting flux qubit\cite{8,9,10,11}
and solid charge qubit\cite{3,4,12}, including single and double dot qubit.
Among them, the gate voltage controlled semiconductor charge qubit has the
potential advantages of being arbitrarily scalable to large system and
compatible with the current microelectronics technology. Recently, Hayashi
et al.\cite{13} have successfully realized coherent manipulation of
electronic state in double-dot system implemented in a GaAs/AlGaAs
heterostructure containing a two dimensional electron gas. The damped
oscillation of population inversion is observed in the time domain and the
dependent of decoherence rate $T_{2}^{-1}$ on the energy offset $\varepsilon
$ is presented. In another similar experiment\cite{14}, the base material
used of the charge qubit is an industry-standard silicon-on-insulator wafer
with aphosphorous-doped active region and all operations (initialization,
manipulation, and measurement) are achieved by capacitively coupled
elements. The change in gate voltage V$_{g3}$ can principally control static
bias $\varepsilon .$ From above, we can see it is necessary to investigate
the effect of the static bias on the tunneling current of the charge qubit.

In this work, we study coherence dynamics of double QDs charge qubit using
spin boson model with static bias, which is investigated without applying
the Markov approximation to the electron-phonon interaction. A simple
explicit expression of population inversion or tunneling current is
presented through perturbation treatment based on unitary transformations.
The coherence-incoherence transition critical point $\alpha _{c}$ versus
bias is gained. Furthermore, the piezoelectric potential phonons induced
decoherence are investigated in detail and possible means for maintaining
quantum coherence are expressed.

The paper is organized as follows: in Sec. 2 we introduce the model
Hamiltonian for static bias spin-boson model and solve it in terms of a
perturbation treatment based on unitary transformations. We analyze the
result and provide proposition for how to maintain tunneling current in Sec.
3. Finally, the conclusion is given in Sec. 4.

2. The model and theory

The model we study is the spin-boson Hamiltonian for static bias\cite{15}
where the two state system is linearly coupled to a continuum of harmonic
oscillators, i.e:
\begin{equation}
H=H_{s}+H_{b}+H_{i}  \label{eq1}
\end{equation}%
where $H_{s}$ is the Hamiltonian of the system, $H_{b}$ of the bath and $%
H_{i} $ of their interaction that is responsible for decoherence.\bigskip\
Here%
\begin{equation}
H_{s}=-\frac{\Delta \sigma _{x}}{2}+\frac{\varepsilon \sigma _{z}}{2}
\label{eq2}
\end{equation}%
\noindent

\begin{equation}
H_{b}=\sum_{k}\omega _{k}b_{k}^{+}b_{k}  \label{eq3}
\end{equation}%
\noindent
\begin{equation}
H_{i}=\frac{1}{2}\sum_{k}g_{k}(b_{k}^{+}+b_{k})\sigma _{z}  \label{eq4}
\end{equation}%
\noindent \noindent with $\sigma _{i}$ being pauli spin matrices. $\Delta $
describes the tunneling coupling between the two states while $\varepsilon $
is the energy offset between the uncoupled charge states. $b_{k}^{+}(b_{k})$
and $\omega _{k}$ are the creation (annihilation) operator and energy of the
phonons with the wave vector k. $g_{k}$ describe the electron-phonon
coupling strength. In this work we consider the static bias case\ with
temperature T=0. The effect of the phonon bath are fully described by a
spectral density:%
\begin{equation}
J(\omega )=\sum_{k}g_{k}^{_{2}}\delta (\omega -\omega _{k}).  \label{eq5}
\end{equation}

In order that the Hamiltonian is diagonalized in $\sigma _{z}$ direction, we
make a displacement to all boson modes,
\begin{equation}
b_{k}=a_{k}-\frac{g_{k}}{2\omega _{k}}\sigma _{0}
\end{equation}%
where $\sigma _{0}$ is a constant and will be determined later. Then we
apply a canonical transformation, $H^{\prime }=\exp (s)H\exp (-s)$ with the
generator\cite{16,17}

\begin{equation}
S=\sum_{k}\frac{g_{k}}{2\omega _{k}}(b_{k}^{+}-b_{k})(\sigma _{z}-\sigma
_{0}).
\end{equation}%
Here we introduce in $S$ a $k$\textit{-}dependent function $\xi _{k}.$ The
aim of the transformation is to take into the correlation between the two
states and the bath. Thus we get the Hamiltonian $H^{\prime }$ and decompose
it into%
\begin{equation}
H^{^{\prime }}=H_{0}^{^{\prime }}+H_{1}^{^{\prime }}+H_{2}^{^{\prime }}
\end{equation}%
where
\begin{equation}
H_{0}^{^{\prime }}=-\frac{1}{2}\eta \Delta \sigma _{x}+\frac{\varepsilon
^{^{\prime }}\sigma _{z}}{2}+\sum_{k}\omega _{k}b_{k}^{+}b_{k}-\sum_{k}\frac{%
g_{k}^{2}}{4\omega _{k}}\xi _{k}(2-\xi _{k})+\sum_{k}\frac{g_{k}^{2}}{%
4\omega _{k}}\sigma _{0}^{2}(1-\xi _{k})^{2}
\end{equation}

\bigskip
\begin{equation}
H_{1}^{^{\prime }}=\frac{1}{2}\sum_{k}g_{k}(1-\xi
_{k})(b_{k}^{+}+b_{k})(\sigma _{z}-\sigma _{0})-i\frac{\eta \Delta }{2}%
\sigma _{y}\sum_{k}\frac{g_{k}}{\omega _{k}}\xi _{k}(b_{k}^{+}-b_{k})
\end{equation}

\bigskip

\begin{eqnarray}
H_{2}^{^{\prime }} &=&-\frac{\Delta \sigma _{x}}{2}(\cosh (\sum_{k}\frac{%
g_{k}}{\omega _{k}}\xi _{k}(b_{k}^{+}-b_{k})-\eta )  \notag \\
&&-i\frac{\Delta \sigma _{y}}{2}(\sinh (\sum_{k}\frac{g_{k}}{\omega _{k}}\xi
_{k}(b_{k}^{+}-b_{k}))-\eta \sum_{k}\frac{g_{k}}{\omega _{k}}\xi
_{k}(b_{k}^{+}-b_{k}))
\end{eqnarray}%
with

\begin{equation}
\eta =\exp (-\sum_{k}\frac{g_{k}^{2}}{2\omega _{k}^{2}}\xi _{k}^{2})
\end{equation}%
and

\begin{equation}
\varepsilon ^{^{\prime }}=\varepsilon -\tau \sigma _{0},\tau =\sum_{k}\frac{%
g_{k}^{2}}{\omega _{k}}(1-\xi _{k})^{2}
\end{equation}%
Obviously, $H_{0}^{^{\prime }}$ can be solved exactly because in which the
spin and bosons are decoupled. Then we can diagonalized $H_{0}^{^{\prime }}$
by a unitary matrix $U,$

\begin{equation}
U=\left(
\begin{array}{cc}
u & v \\
v & -u%
\end{array}%
\right) ,
\end{equation}%
with%
\begin{equation}
u=\frac{1}{\sqrt{2}}(1-\sin \theta )^{\frac{1}{2}},v=\frac{1}{\sqrt{2}}%
(1+\sin \theta )^{\frac{1}{2}}
\end{equation}%
and $\sin \theta =\varepsilon ^{^{\prime }}/W$ with $W=(\varepsilon
^{^{\prime }2}+\eta ^{2}\Delta ^{2})^{1/2}.$

$H^{^{\prime }}$ is transformed as follows (to the second order of $g_{k}$):%
\begin{equation}
H^{^{\prime \prime }}=U^{+}H^{^{\prime }}U=H_{0}^{^{\prime \prime
}}+H_{1}^{^{\prime \prime }}+H_{2}^{^{\prime \prime }}
\end{equation}

\begin{equation}
H_{0}^{^{\prime \prime }}=-\frac{1}{2}W\sigma _{z}+\sum_{k}\omega
_{k}b_{k}^{+}b_{k}-\sum_{k}\frac{g_{k}^{2}}{4\omega _{k}}\xi _{k}(2-\xi
_{k})+\sum_{k}\frac{g_{k}^{2}}{4\omega _{k}}\sigma _{0}^{2}(1-\xi _{k})^{2}
\end{equation}

\begin{eqnarray}
H_{1}^{^{\prime \prime }} &=&-\frac{1}{2}\sum_{k}g_{k}(1-\xi
_{k})(b_{k}^{+}+b_{k})(\frac{\varepsilon }{W}\sigma _{z}+\sigma _{0})  \notag
\\
&&+\frac{\eta \Delta }{2W}\sigma _{x}\sum_{k}g_{k}(1-\xi
_{k})(b_{k}^{+}+b_{k})+i\frac{\eta \Delta }{2}\sigma _{y}\sum_{k}\frac{g_{k}%
}{\omega _{k}}\xi _{k}(b_{k}^{+}-b_{k})
\end{eqnarray}%
$H_{1}^{^{\prime \prime }}$ and $H_{2}^{^{\prime \prime }}$ are treated as
perturbation and they should be as small as possible. For this purpose $%
\sigma _{0}$ and $\xi _{k}$ are determined in such a way
\begin{equation}
\sigma _{0}=-\frac{\varepsilon }{W};\xi _{k}=\frac{\omega _{k}}{\omega _{k}+W%
}
\end{equation}%
\ that%
\begin{equation}
H_{1}^{^{\prime \prime }}=\frac{\eta \Delta }{2}\sum_{k}\frac{g_{k}\xi _{k}}{%
\omega _{k}}\left[ b_{k}^{+}(\sigma _{x}+i\sigma _{y})+b_{k}(\sigma
_{x}-i\sigma _{y})\right]
\end{equation}%
and $H_{1}^{^{\prime \prime }}|g\rangle =0.$ This is the key point in our
approach. We should remark that
\begin{equation}
\varepsilon ^{^{\prime }}=\varepsilon (1+\frac{\tau }{W});\sin \theta =\frac{%
\varepsilon (1+\frac{\tau }{W})}{W}.
\end{equation}%
So $\theta $ is in the range of $0\leq \theta \leq \pi /2$ and the sign of $%
\sin \theta $ is identical with static bias. If $\theta =0$, the model
correspond to zero bias case.

We denote the ground state of $H_{0}^{^{\prime \prime }}$ as $|g\rangle
=|s_{1}\rangle |\{0_{k}\}\rangle $, and the lowest excited states as $%
|s_{2}\rangle |\{0_{k}\}\rangle $, $|s_{1}\rangle |\{1_{k}\}\rangle $ where $%
|s_{1}\rangle ,$ $|s_{2}\rangle $ are eigenstates of $\sigma _{z}$ ($\sigma
_{z}|s_{1}\rangle =|s_{1}\rangle ,$ $\sigma _{z}|s_{2}\rangle
=-|s_{2}\rangle $), $|\{n_{k}\}\rangle $ means that there are $n_{k}$
phonons for mode k. Thus, we can diagonalize $H^{^{\prime \prime }}$ as:%
\begin{equation}
H^{^{\prime \prime }}=-\frac{1}{2}W|g\rangle \langle g|+\sum_{E}E|E\rangle
\langle E|+\text{terms with higher excited states}
\end{equation}%
the experiments in Ref.13,14 are performed at lattice temperature below or
about 20 mK. At such a low temperature, the multiphoton process is weak
enough to be negligible. So we can diagonalize through the following
transformation\cite{16}%
\begin{equation}
|s_{2}\rangle |\{0_{k}\}\rangle =\sum_{E}x(E)|E\rangle
\end{equation}%
\begin{equation}
|s_{1}\rangle |\{1_{k}\}\rangle =\sum_{E}y_{k}(E)|E\rangle
\end{equation}

\begin{equation}
|E\rangle =x(E)|s_{2}\rangle |\{0_{k}\}\rangle
+\sum_{k}y_{k}(E)|s_{1}\rangle |\{1_{k}\}\rangle
\end{equation}%
where
\begin{equation}
x(E)=(1+\sum_{k}\frac{V_{k}^{2}}{(E+\frac{1}{2}W-\omega _{k})^{2}})^{\frac{1%
}{2}}
\end{equation}

\bigskip
\begin{equation}
y_{k}(E)=\frac{V_{k}}{E+\frac{1}{2}W-\omega _{k}}x(E)
\end{equation}%
with $V_{k}=\eta \Delta g_{k}\xi _{k}/\omega _{k}.$ E are the diagnalized
excitation energy and they are solutions of the equation
\begin{equation}
E-\frac{1}{2}W-\sum_{k}\frac{V_{k}^{2}}{E+\frac{1}{2}W-\omega _{k}}=0
\end{equation}%
So Hamiltonian can approximately be describes as:%
\begin{equation}
H^{^{\prime \prime }}=-\frac{1}{2}W|g\rangle \langle g|+\sum_{E}E|E\rangle
\langle E|
\end{equation}

The population inversion can be defined as $p(t)=\left\langle \psi
(t)\left\vert \sigma _{z}\right\vert \psi (t)\right\rangle $, where $%
\left\vert \psi (t)\right\rangle $ is the total wave function in the
Schrodinger picture. Since the initialization of the charge qubit is used to
in the state $\left\vert L\right\rangle ,$ it is reasonable to choose
initial state $\left\vert \psi (0)\right\rangle =e^{-s}\left\vert
L\right\rangle \left\vert 0_{k}\right\rangle .$ Then we can obtain%
\begin{eqnarray}
p(t) &=&\left\langle \psi (0)\left\vert Ue^{iH^{^{\prime \prime
}}t}U^{+}e^{s}UU^{+}\sigma _{z}UU^{+}e^{s}Ue^{-iH^{^{\prime \prime
}}t}U^{+}\right\vert \psi (0)\right\rangle \\
&=&-u^{2}\sin \theta -v^{2}\sin \theta \sum_{kk^{^{\prime }}EE^{^{\prime
}}}y_{k^{^{\prime }}}^{\ast }(E^{^{\prime }})y_{k}(E)x(E)x^{\ast
}(E)e^{i(E-E^{^{\prime }})t}  \notag \\
&&-v^{2}\sin \theta \sum_{EE^{^{\prime }}}\left\vert x(E)\right\vert
^{2}\left\vert x(E^{^{\prime }})\right\vert ^{2}e^{i(E-E^{^{\prime }})t}
\notag \\
&&+uv\cos \theta \sum_{E}(\left\vert x(E)\right\vert ^{2}e^{i(E+\frac{w}{2}%
)t}+\left\vert x(E)\right\vert ^{2}e^{-i(E+\frac{w}{2})t})
\end{eqnarray}%
For the different terms appearing in Eq.(31), we employ the orthogonal
property%
\begin{equation}
\sum_{k}y_{k}(E)y_{k}(E^{^{\prime }})=\delta (E-E^{^{\prime
}})-x(E)x(E^{^{\prime }})
\end{equation}%
and the residue theorem%
\begin{eqnarray}
\sum_{E} &\mid &x(E)\mid ^{2}e^{iEt} \\
&=&e^{-i\frac{W}{2}t}\frac{1}{2\pi i}\int \frac{e^{i(E+\frac{W}{2})t}dE}{(E-%
\frac{1}{2}W-\sum_{k}\frac{V_{k}^{2}}{E+\frac{1}{2}W-\omega _{k}})} \\
&=&e^{-i\frac{W}{2}t}\frac{1}{2\pi i}\int \frac{e^{iE^{^{\prime
}}t}dE^{^{\prime }}}{(E^{^{\prime }}-W-\sum_{k}\frac{V_{k}^{2}}{E^{^{\prime
}}-\omega _{k}})}
\end{eqnarray}%
Denoting the real and imaginary part of $\sum_{k}\frac{V_{k}^{2}}{\omega
-\omega _{k}\pm i0^{+}}$ as R($\omega $) and $\mp \gamma (\omega )$,
respectively, we can get
\begin{eqnarray}
R(\omega ) &=&\wp \sum_{k}\frac{V_{k}^{2}}{\omega -\omega _{k}} \\
&=&(\eta \Delta )^{2}\wp \int\limits_{0}^{\infty }d\omega ^{^{\prime }}\frac{%
J(\omega ^{^{\prime }})}{(\omega -\omega ^{^{\prime }})(\omega ^{^{\prime
}}+\eta \Delta )^{2}}
\end{eqnarray}

\begin{equation}
\gamma (\omega )=\pi \sum_{k}V_{k}^{2}\delta (\omega -\omega _{k})=\pi (\eta
\Delta )^{2}\frac{J(\omega )}{(\omega +\eta \Delta )^{2}}
\end{equation}%
where $\wp $ stands for Cauchy principal value, and $J(\omega )$ is the
spectral density.

The contour integral in Eq.(35) can be proceed by calculating the residue of
integrand and substituting in Eq.(31), we obtain%
\begin{equation}
p(t)=-\sin \theta +\sin \theta (1+\sin \theta )e^{-2\gamma (\omega )t}+\cos
^{2}\theta \cos (\omega _{0}t)e^{-\gamma (\omega )t}
\end{equation}%
where $\omega _{0}$ is the solution to the equation
\begin{equation}
\omega -W-R(\omega )=0
\end{equation}%
Thus a rather simple expression for the dynamical tunneling is obtained
analytically. It should be noted here that for t$\rightarrow \infty ,$ our
result tend towards thermodynamics equilibrium value\cite{18}, which is
modulated by static bias%
\begin{equation}
p(\infty )=-\sin \theta =-\frac{\varepsilon (1+\frac{\tau }{W})}{W}
\end{equation}

Comparing with Markovian approximation, our result not only give the long
time limit behavior, but the short time damping coherence oscillation. That
must holds most prominently if semiconductor quantum dots are to be used as
basic building blocks for quantum information processing where the operation
completely relies on the presence of coherence.

Finally, the tunneling electron population in the right dot at time t can be
obtain n(t)=$\frac{1+\left\langle \sigma _{z}(t)\right\rangle }{2}=\frac{%
1+p(t)}{2}.$

Until here, our presentation is not restricted by the form of the spectral
density and can be extended to all kinds of baths. Previous work states
that, at zero temperature in double-dot system of GaAs material, the
dominant contribution of phonons to QDs come from the piezoelectric coupling
and the deformation potential coupling is small enough to be ignored. So we
will use the piezoelectric coupling spectral density\cite{17,19}:

\begin{equation}
J(\omega )=\alpha \omega (1-\frac{\omega _{d}}{\omega }\sin \frac{\omega }{%
\omega _{d}})\theta (\omega _{c}-\omega )
\end{equation}%
where $\alpha $ is the dimensionless coupling constant, $\omega _{c}=s/l$
and $\omega _{d}=s/d$ (s is the sound velocity in crystal, l is the dot size
and d is the center-to-center distance between two dots) and $\theta (x)$ is
the usual step function.

3. The result and discussion

From Eq.(39) we have been able to obtain the population inversion of charge
qubit $p(t)$ as the analytical damped oscillation form with frequency $%
\omega _{0}$ and damping rate $\gamma (\omega ).$ In the following
calculations, $\omega _{c}$ is taken as the energy unit. We choose
the quantum dot size l as 100nm (approximate size for the dot in
Ref.13), i.e. $ \omega _{c}=32.5$ $\mu $eV (or $0.05 $ $\rm
ps^{-1}$). Assume the distance between two dots is sufficiently
large, d=667 nm, correspondingly $\omega
_{d}=0.15\omega _{c}$. The typical value of tunneling barriers in experiment%
\cite{13} are $\Delta =9$ $\mu $eV. Without special indication, $
\Delta $ is choosen 9 $ \mu$eV.

First we illustrate the population inversion as function of time
($\omega _{c}t$) in fig.1(a) with five different static bias:
$\varepsilon =-8$ $\mu $eV (dashed line), $0 \mu $eV(solid line), $8
\mu $eV (dotted line) for fixed $\alpha =0.04$. It is clearly shown
that the population inversion exhibit damping oscillation and
symmetry broken. The asymptotic value of t he long time limit
$p(\infty )$ is determined by the static bias. If bias is zero,
$p(\infty )=0.$

Fig.1(b) presents the long time limit of population inversion or the
thermodynamics equilibrium value $p(\infty )=-\varepsilon ^{^{\prime }}/W$
as functions of bias $\varepsilon .$ Comparing with the results of NIBA\cite%
{18}, which is expressed as $p(\infty )=-\varepsilon /\sqrt{\varepsilon
^{2}+\Delta _{r}^{2}},$ on the scale of the figures, the two curves are
overlapped if the identical Ohmic spectral density $J(\omega )=2\alpha
\omega \theta (\omega _{c}-\omega )$ is used. While the system is in the
piezoelectric potential bath, $p(\infty )$ deviate a little.

Fig.1(c) sketches the coherence-incoherence critical point $\alpha
_{c}$ versus $\varepsilon /\omega _{c}$ for fixed $\Delta =9$ $\mu
eV$ in Ohmic bath. We see that bias increases $\alpha _{c}$
significantly and induce a transition from strongly damped
incoherent to coherent oscillation behavior. The similar result is
also obtained by Klaus\cite{15} from Monte Carlo simulations.

The quantum coherence depends on two factors: oscillation frequency $\omega
_{0}$ and decoherence rate $T_{2}^{-1}.$ It is already noticed that long
coherence time is favorable for quantum manipulation. However low
oscillation frequency means that the number of quantum operation that can be
achieved within the coherence time is very limit\cite{14}. So we must give
attention to two aspects coherence time and oscillation frequency. In what
follows, we analyse the effect of the static bias $\varepsilon ,$ the
tunneling energy $\Delta $ and e-p coupling constant $\alpha $ on the
decoherence rate $\gamma $ and the damping oscillation frequency $\omega $
and elucidate the advantage and disadvantage for changing characteristic
energy $\Delta $ and $\varepsilon $ or decreasing e-p coupling to remain
quantum coherence.

In contrast with experiment, we present the decoherence rate $\gamma $ ($%
T_{2}^{-1}$) as function of energy offset $\varepsilon $[Fig. 2]. The
electron-phonon coupling constant $\alpha =0.02-0.07$ is used to explain the
inelastic current in GaAs/AlGaAs heterostructure DQD samples\cite{19}. In
Fig.2(a) we shows the damping rate $\gamma $ as a function of static bias $%
\varepsilon $ with three different e-p coupling $\alpha =0.04$ (solid), $%
\alpha =0.08$ (dashed) $\alpha =0.12$ (dotted). These curves show
clearly that the decoherence rate raise with increasing of the
absolute value of the bias $\varepsilon $. The vary of $\gamma $ vs
$\varepsilon $ is agree with the result of experiment. If
$\varepsilon =0$ $\mu$eV and $\alpha =0.04,$ decoherence rate
$\gamma $ approximately is 0.2 $\rm  ns^{-1}$. Increasing the e-p
coupling to $\alpha =0.12$ the damping rate contributing from
piezoelectric phonons grow up to 0.50 $\rm  ns^{-1}$, occupy more
than half of the experimental result. It proves that the coupling to
the piezoelectric potential phonons is one of the main decoherence
mechanisms in such a double-dot system.

The damping oscillation frequency $\omega _{0}$ vs bias $\varepsilon
$ is shows in Fig.2(b), and the parameters are same as those in
Fig.2(a). The qubit oscillation frequency can be changed continually
by bias voltage and behave nonlinear response versus $\varepsilon ,$
which explain the experimental result in Ref.14. Setting large
static bias can efficiently improve oscillation frequency. When
$\alpha =0.04$ and $\varepsilon =0$ $\mu$eV, the frequency
approximately is 2.1 GHz, which is extraordinarily agree with the
fit result of the experimental data in Ref.15. The decoherence rate
and the oscillation frequency reveal symmetry about bias.

Fig.3(a) presents the decoherence rate as function of tunneling coupling $%
\Delta $ at three different static bias $\varepsilon =0$ $\mu$eV
(solid), $3$ $\mu $eV (dashed), $6$ $\mu $eV (dotted)$,$ fixing
$\alpha $ as 0.04. Enlarging tunneling coupling $\Delta $ between
two dots make damping rate $\gamma $ gradually increase as shown in
Fig.3(a). So quantum coherence can be remained by reducing tunneling
barrier. In experiment, the barrier tunneling can be principally
defined by the materials and the geometrical constriction between
the dots in the fabrication of the device. But it is possible to
modify the tunneling barrier by gate voltages, just as shown in
Ref.13 Fig.2(d). From Fig.3(a), if $\Delta $ in the order of
magnitude of $\mu $eV, the decoherence rate is probably in the range
of $0.2\sim 1$ $\rm ns^{-1}$, which is agree well
with the Ref.13. A plot of oscillation frequency $\omega $ as function $%
\Delta $ is shown in Fig.3(b). At zero bias $\varepsilon =0,$ the
curve behavior is approximately line. Increasing static bias,
$\omega $ appear nonlinear response. Comparing two similar
experiment in Ref.15 and Ref.16, the characterized energies $E^{\ast
}$ 40.5 neV corresponding to angular frequency $\omega =62$ MHz in
the silicon two-level quantum system are 1/37 smaller than quantum
level spacing 1.5 $\mu $eV corresponding to oscillation frequency
$\omega =2.3$ GHz in the reports for semiconductor double quantum
dot charge qubit. From the calculation result, we can deviate that
the reducing of energy split make damping rate and oscillation
frequency fall same order of magnitude, correspondingly, in the
piezoelectric coupling spectral density, so the lack of
piezoelectric coupling in the silicon QD might result in the long
coherence time two orders of magnitude longer than reports for
semiconductor QD. We find that the bias can effectively enhance
oscillation frequency but have relative little lifting for damping
rate in the case of weak tunneling couple. Therefore adjust static
bias larger is another choice to preserve coherence and that is a
suggestion to overcome the obstacle in Ref.16.

Fig.4 presents the decoherence rates $\gamma $ and oscillation frequency $%
\omega $ as functions of dimensionless coupling constant $\alpha $
at three different bias $\varepsilon =0$ $\mu $eV (solid), $3$ $\mu
$eV (dashed), $6$ $\mu $eV (dotted)$.$ As shown in Fig.4(a), for
determined bias, it is an almost linear relation between damping rate $%
\gamma $ and dimensionless coupling constant $\alpha $. Fig.4(b) displays
the oscillation frequency $\omega $ versus coupling constat $\alpha $. These
curves show clearly that when the coupling constant $\alpha $ become weaker
three or two orders of magnitude from 0.04, the oscillation frequency $%
\omega $ have linearly increased, albeit very relaxedly. Applied static
bias, the decoherence rate and oscillation frequency always are enhanced. As
a result, to keep the coherence of charge qubit in long time, one good way
is to minish the coupling with environment, either by find good material
with small e-p coupling, or by modifying the design of the DQD structure, or
by better gate tuning.

Anyway, to realize quantum computation, the important direction are to
reduce the coupling to environment by all means, but to ensure effective
interdot coupling and static bias.

4. Conclusion

We studied the charge qubit dynamics with static bias by a perturbation
treatment based on two times unitary transformations. Our approach applies
to all forms of spectral density. In Ohmic bath, the result shows that $%
\alpha _{c}$ clearly increases and population inversion breaks symmetry in
biased system. Analyzing the piezoelectric coupling phonon induced
decoherence, we find that, weak coupling of the charge qubit to the
environment carries out large coherence oscillation frequency and long
coherence time. When tunneling coupling between two QDs is fixed, adjust
gate voltage to enlarge static bias make oscillation frequency observably
increase while damping rate unnoticeably increase. That is another better
choice to maintain quantum coherence. Finally we hope that our predictions
can offer advice for experimenter and be testified by experiment in the near
future.

Acknowledgments: This work was supported by the China National Natural
Science Foundation (Grants No. 10474062 and No. 90503007).

--------------------

\newpage

\begin{description}
\item {\large FIGURES}

\vspace{0.3cm} 
\vspace{0.3cm} 

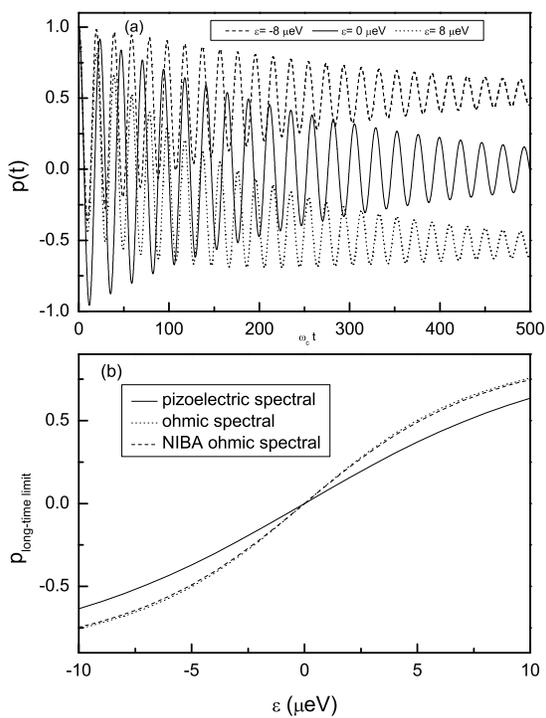
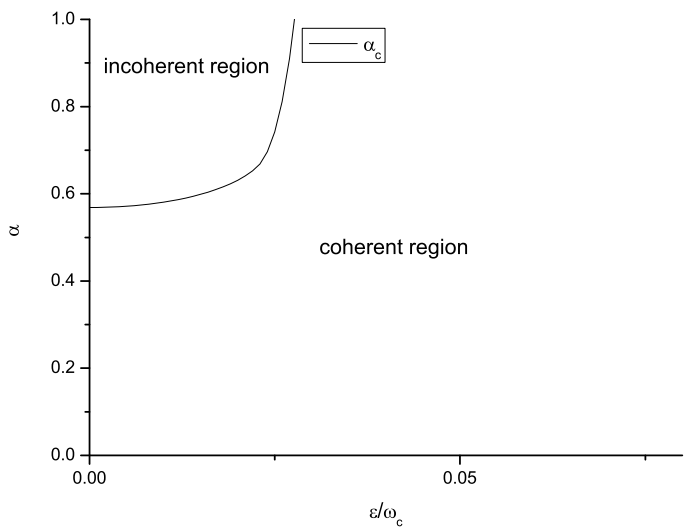
\item {Fig.1.(a) The population difference as a function of }$\omega _{c}t$
for different static bias $-8$ $\mu $eV (dashed line)$,0$ $\mu $eV
(solid line), $8$ $\mu $eV (dotted line). (b) The long time limit
$p(\infty )$ (solid) as a function of static bias $\varepsilon $ in
piezoelectric potential bath. $p(\infty )$ (dotted) in Ohmic bath and $%
p(\infty )$ (dashed) in Ohmic bath for NIBA. The coupling constant
is fixed as $\alpha =0.04.$ (c) The coherence-incoherence critical
point $\alpha _{c}$ versus static bias $\varepsilon /\omega _{c}$
with $\Delta =9$ $\mu eV $.


\vspace{0.3cm} 
\vspace{0.3cm} 

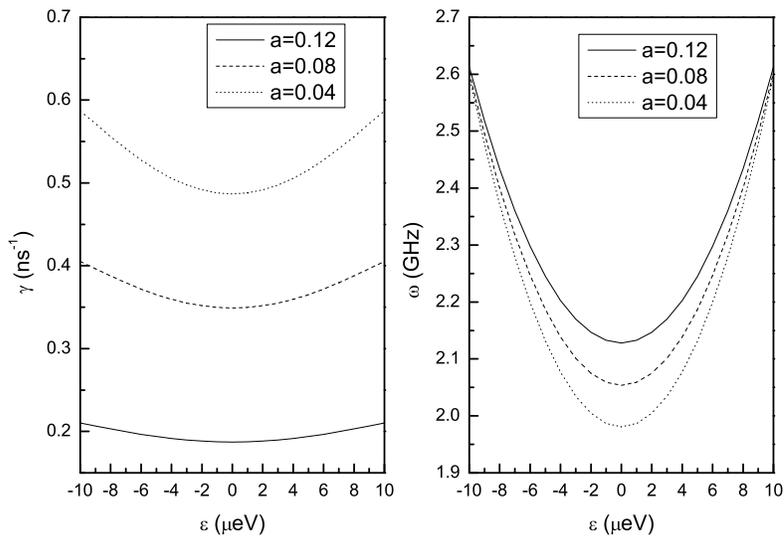
\item {Fig.2. The} damping rate $\gamma $ and {oscillation frequency }$%
\omega $ {versus }$\varepsilon $ in (a) and (b), respectively with
$\Delta =9 $ $\mu $eV, $\omega _{d}=0.15\omega _{c}$ for different
$\alpha =0.04$ (solid line), $0.08$ (dashed line), $0.12$ (dotted
line).


\vspace{0.3cm} 
\vspace{0.3cm} 

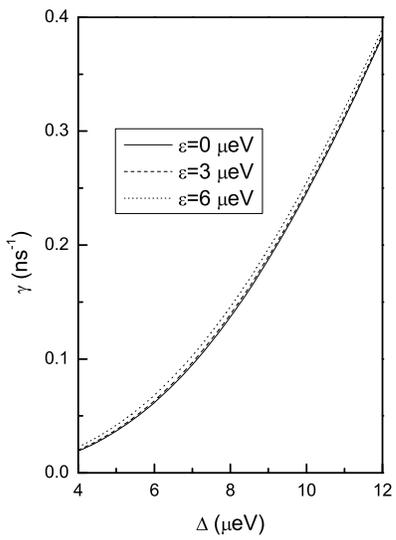
\item {Fig.3} {\ The} damping rate $\gamma $ and {oscillation frequency }$%
\omega $ {versus }$\Delta $ in (a) and (b), respectively with $\alpha =0.04$%
, $\omega _{d}=0.15\omega _{c}$ for different static bias $\varepsilon =0$ $%
\mu $eV (solid line), $3$ $\mu $eV (dashed line), $6$ $\mu $eV
(dotted line).


\vspace{0.3cm} 
\vspace{0.3cm} 

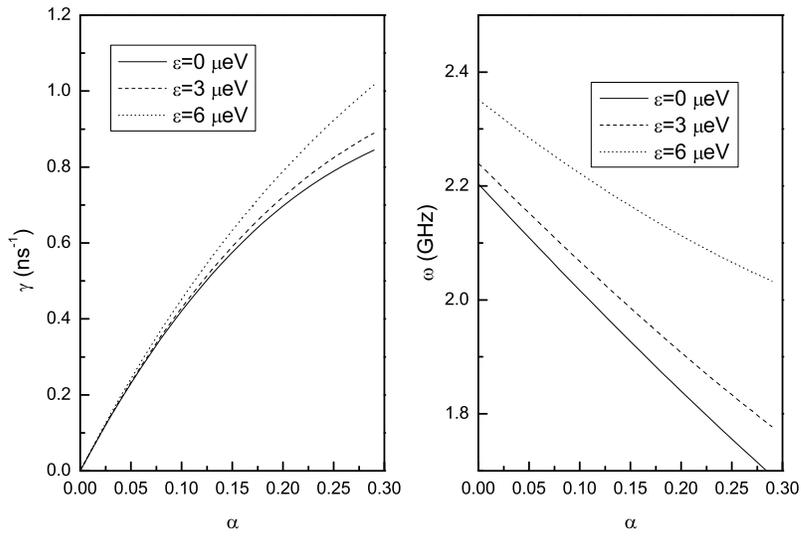
\item {Fig.4. The } damping rate $\gamma $ and {oscillation frequency }$%
\omega $ {versus }$\alpha $ in (a) and (b), respectively with $\Delta =9$ $%
\mu $eV, $\omega _{d}=0.15\omega _{c}$ for different static bias $%
\varepsilon =0$ $\mu $eV (solid line), $3$ $\mu $eV (dashed line),
$6$ $\mu $eV (dotted line).


\vspace{0.3cm} 
\vspace{0.3cm} 
\end{description}


\begin{thebibliography}{100}


\bibitem{1} M. A. Nielsen and I. L. Chuang, \textit{Quantum computation and
quantum information} (Cambridge University Press, United Kingdom, 2000).

\bibitem{2} S. Tarucha, D. G. Austing and T. Honda, Phys. Rev. Lett. \textbf{%
77}, 3613 (1998).

\bibitem{3} T. H. Oosterkamp \textit{et al}., Nature \textbf{395,} 873
(1998).

\bibitem{4} Toshimasa Fujisawa, Tjerk. H. Oosterkamp \textit{et al}.,
Science \textbf{282}, 932 (2001).

\bibitem{5} V. N. Stavrou and Xuedong Hu, Phys. Rev. B \textbf{72}, 075362
(1999).

\bibitem{6} W. G. Unruh, Phys. Rev. A \textbf{51}, 992 (1995).

\bibitem{7} S. D. Barrett, G. J. Milburn, Phys. Rev. B \textbf{68}, 155307
(2003).

\bibitem{8} P. Bertet, I. Chiorescu, G. Burkard \textit{et al.}, Phys. Rev.
Lett. \textbf{95}, 257002 (2005).

\bibitem{9} I. Chiorescu, P. Bertet, K. Semba \textit{et al.}, Nature
\textbf{431,} 159 (2004).

\bibitem{10} I. Chiorescu, Y. Nakamura, C. J. P. M. Harmans and J. E. Mooij,
Science \textbf{299}, 1869 (2003).

\bibitem{11} A. Lupascu, E. F. C. Driessen. L. Roschier \textit{et al.},
Phys. Rev. Lett. \textbf{96}, 127003 (2006).

\bibitem{12} S. Gardelis, C. G. Smith and J. Cooper \textit{et al.}, Phys.
Rev. B \textbf{67}, 073302 (2003).

\bibitem{13} T. Hayashi et al, Phys. Rev. Lett. \textbf{91}, 226804 (2003).

\bibitem{14} J. Gorman, D. G. Hasko and D. A. Williams, Phys. Rev. Lett.
\textbf{95}, 090502 (2005).

\bibitem{15} Klaus Volker, Phys. Rev. B \textbf{58}, 1862 (1998).

\bibitem{16} H. Zheng, Eur. Phys. J. B \textbf{38}, 559 (2004).

\bibitem{17} Zhuo-Jie Wu, Ka-Di Zhu \textit{et al.}, Phys. Rev. B \textbf{71}%
, 205323 (2005).

\bibitem{18} U. Weiss, \textit{Quantum dissipative system},\textit{\ }(World
Scientific, Singapore, 1993).

\bibitem{19} T. Brandes and T. Vorranth, Phys. Rev. B \textbf{66}, 075341
(2002).
\end{thebibliography}
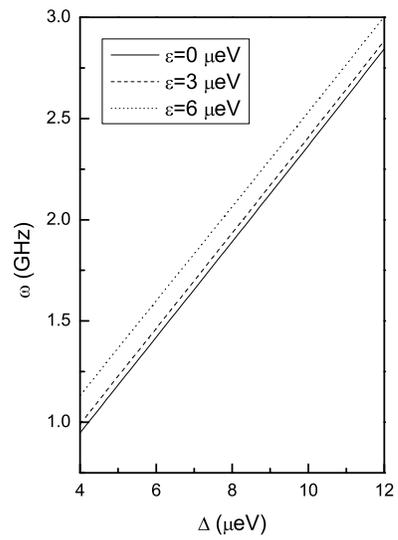
\end{document}